\documentstyle[floats,prd,aps,epsf]{revtex} 
\draft % command makes pacs numbers print 

\begin{document}

\twocolumn[\hsize\textwidth\columnwidth\hsize\csname
@twocolumnfalse\endcsname

\def\boxit#1{\vbox{\noindent\hrule\hbox{\vrule\hskip 3pt
\vbox{\vskip 3pt
\hbox{#1}\vskip 3pt}\hskip 3pt\vrule}\hrule}}

\overfullrule 0pt

\title{Brief Comments on ``The Shapiro Conjecture, Prompt
or Delayed Collapse ?'' by Miller, Suen and Tobias}

\author{Stuart L. Shapiro}

\address{Departments of Physics and Astronomy, \& NCSA, University of Illinois at 
Urbana-Champaign, Urbana, IL 61801}

\maketitle

\begin{abstract} 
Recent numerical simulations address a conjecture
by Shapiro that when two neutron stars collide head-on from rest
at infinity, sufficient thermal pressure may be generated to support
the hot remnant in quasi-static equilibrium against collapse prior to
neutrino cooling. The conjecture is meant to apply even when the total 
remnant mass exceeds the maximum mass of a cold neutron star.
One set of simulations seems to corroborate the conjecture, while another,
involving higher mass progenitors each very close to the maximum  mass, does not. 
In both cases the total mass of the remnant exceeds the maximum mass. 
We point out numerical subtleties in performing such simulations when 
the progenitors are near the maximum mass; they can explain why  
the simulations might have difficulty assessing the conjecture in such
high-mass cases.

\end{abstract}

\vskip2pc]

Shapiro~\cite{SLS} recently speculated that following the head-on
collision of two neutron stars from rest at infinity, the two recoil
shocks which propagate back into each star following contact
might generate sufficient thermal pressure to hold up the remnant 
against gravitational collapse, at least until it cools via neutrino emission.
This quasi-equilibrium state will last many dynamical
timescales ($\sim \mbox{msec }$) because the neutrinos,
which eventually carry off the thermal energy, leak out slowly, on the
neutrino diffusion timescale ($\sim 10~\mbox{sec}$).
The argument  was independent of the total mass of the progenitors and 
applies even if the mass of the remnant greatly exceeds the maximum mass of a
cold neutron star. A simple, analytic analysis using
relativistic polytropes served to motivate the conjecture. 
The analysis assumed strict
conservation of rest-mass and total mass-energy in the collision,
as well as relaxation, following merger, to the same polytropic 
density profile in the 
remnant as in the progenitors.  Given these assumptions it was possible to
show that a stable equilibrium solution always
exists for the remnant, independent of mass. In evaluating 
these assumptions, Shapiro argued that while they 
greatly oversimplify the collision in order to permit an analytic
proof, they do provide a reasonable approximate description of
the idealized head-on scenario under consideration. He did show that
the loss of energy due to gravitational radiation, though small, would
rule out an equilibrium solution for configurations arbitrarily close to
the maximum mass, if the collision were not accompanied by loss of
rest-mass. More significantly, 
he cautioned that a dynamical collision need not
relax to a stable, equilibrium solution after all, even though 
one exists, but could instead 
overshoot the equilibrium state and collapse to black hole.
But he put forward the possibility of delayed collapse 
as a plausible outcome of head-on collisions from rest at infinity
and one which might be realized in numerical simulations.

Numerical simulations to test this conjecture have been reported
recently in Ref~\cite{MST}.  The simulations treat colliding $\Gamma = 2$ polytropes in a 3+1 numerical scheme
that solves Einstein's field equations of general relativity
coupled to the equations of relativistic hydrodynamics.
The simulations employed $\Gamma = 2$ polytropic
models that satisfy the TOV equilibrium equations for isolated, spherical
stars. The stars are placed at {\it finite} separation, no more 
than $3R$ apart, where $R$ is the stellar radius. These configurations
are then boosted towards each other with the {\it Newtonian} 
freefall velocity~\cite{freefall}. 
No attempt is made to correct for the 
distortions in the initial matter density or pressure profiles that 
arise at such close separation due to tidal interactions. 
The simulations described in Ref ~\cite{MST} deal with collisions of
$1.4 M_{\odot}$ stars ~\cite{mass}. We shall assume below that
this quoted mass represents the total mass-energy of each of the progenitors, 
although no distinction is made in Ref ~ \cite {MST} between 
rest-mass $M_0$ and total mass-energy $M$, nor is there any 
discussion about how the mass in the colliding system 
is actually measured and monitored numerically.
The subsquent evolution of this model produces a black hole 
a few dynamical timescales following contact and merger. This example,
it is claimed in Ref~\cite{MST}, 
is in contradiction to the conjecture of Shapiro.

Two aspects of the numerical simulations are crucial to note.
One is that the maximum stable mass of the polytropic equation of state is 
$1.46 M_{\odot}$ in the units of Ref~\cite{MST}. The 
other, alluded to but not reported in Ref~\cite{MST}, is that a
second set of simulations using the same equation of state for
the head-on collision of $0.8 M_{\odot}$ configurations 
did {\it not} result in collapse following merger~\cite{priv}. 
In this case the 
simulations show that the merged remnant is supported in stable 
equilibrium by the thermal pressure 
generated by collision-induced shock heating, in apparent 
accord with the Shapiro conjecture. 
Interestingly, in both cases the total mass
of the merged remnant exceeds the maximum mass of a cold star.

Does the simulation with $1.4 M_{\odot}$ stars show unambiguously
that a delayed collapse does not occur for stars of this mass?
Insight into what could
be happening may be gleaned from the original discussion 
in Ref~\cite{SLS}. 
Section VI of that paper,  ``Discussion and Caveats '' 
already anticipated numerical subtleties
that would be encountered in simulating a collision  
of stars that are initially at rest at infinity. 
To obtain reliable results, the 
requirements for computational accuracy become
very stringent for high-mass stars near the maximum mass. 
The reason is clear from the analytic argument for the 
existence of a stable, equilibrium solution for the remnant. 
For such a solution to exist, 
it is necessary that the binding energy of the final remnant
reside along the stable branch of the TOV equilibrium curve. 
The ratio $M/M_0$ measures the specific binding energy of a star 
according to $E_{\rm bind}/M_0=1-M/M_0$. This ratio 
monotonically {\it decreases} with increasing
central density $\rho_c$ along the stable branch
of the TOV curve, whereas the mass and 
rest-mass monotonically {\it increase}
along this branch. The turning point on the curve marks the onset of
radial instability; there $M$ and $M_0$ asssume their 
maximum values and the binding
energy ratio assumes its minimum value, $(M/M_0)_{\rm min}$.
In the idealized scenario in which the stars begin from rest at infinity, 
strict conservation of mass and rest-mass imply that for the hot remnant 

\begin{equation}
\label{one}
M_{0,{\rm hot}}\ =\ 2M_0\ ~\mbox{and}~\ M_{\rm hot}\ =\ 2M.
\end{equation}
As a result, the binding
energy of the hot remnant is identical to the binding energy of each of the progenitor stars, and
this guarantees the existence of an equilibrium solution 
(cf. Eqs. 8 - 10 of Ref ~\cite{SLS}).
 
Now consider an alternative scenario in which Eq. ~(\ref{one}) 
is replaced by

\begin{equation}
\label{two}
M_{0,{\rm hot}}\ =\ 2M_0\ ~\mbox{and}~\ M_{\rm hot}\ =\ 2M (1-f).
\end{equation}
As discussed in Ref~\cite{SLS}, the fractional change in the total mass-energy $f$ might be 
the result of a true physical departure from the conditions or assumptions of the idealized
collision scenario. The quantity $f$ may account, for example, for the
energy loss due to gravitational radiation or for
the release of the progenitor stars from rest at 
finite separation rather from infinity~\cite{rad}. Alternatively, $f$ might 
represent the degree of {\it numerical error} arising in a numerical 
simulation. Such error might be
present due to imprecise initial data, including, for example, initial data 
which do not exactly correspond to the true solution at finite 
separation for two stars which actually
begin from rest at infinity~\cite{init}. 
Numerical error might also result from 
finite-difference integration error in the evolution due to finite 
spatial grid and time-step resolution, imprecise outer boundary conditions or
outer boundary points that don't extend sufficiently far into the
radiation zone, global instabilities, etc~\cite{error}.
To illustrate, suppose we assume
that for whatever reason, numerical error results in a spurious fractional
decrease in $M$ in the course of the simulation. We then estimate the
the maximum tolerable error $f_{\rm tol} \geq 0$ which still admits a stable,
equilibrium solution for
the remnant~\cite{tol}. According to Eq. ~(\ref{two}), relaxation to a stable equilibrium
state requires

\begin{equation}
\label{three}
{M_{\rm hot}\over M_{0,{\rm hot}}} = (1-f)  {M \over M_0}
\geq  \left ({M \over M_0}\right )_{\rm min}
{\Longrightarrow  {f_{\rm tol} \leq 1-
{\left  ({M \over M_0}\right )_{\rm min}
\over \left ({M \over M_0}\right )}}},
\end{equation}

Eqn. ~(\ref{three}) demonstrates the difficutly
of simulating accurately a head-on collision
from rest at infinity as the progenitor masses approach the maximum mass.
As $M \rightarrow M_{\rm max}$, we have $(M/M_0) \rightarrow (M/M_0)_{\rm min}$
and, hence $f_{\rm tol} \rightarrow 0$.

To appreciate the severity of the problem, consider the actual 
simulations performed for $\Gamma = 2$, for which
$(M/M_0)_{\rm min} = 0.910$ (see Fig 3 in Ref ~\cite{SLS}). 
This minimum occurs at $1.46 M_{\odot}$, the maximum mass.
We then find 
$(M/M_0) = 0.915$ for the $M=1.4 M_{\odot}$ star
and $(M/M_0) = 0.959$
for the $M=0.8 M_{\odot}$ star. According to Eq. ~(\ref{three}) these
values imply

\begin{equation}
\label{four}
{f_{\rm tol} (0.8 M_{\odot}) = 0.05 ~~~{\rm and} ~~~ f_{\rm tol} (1.4 M_{\odot}) = 0.005.}
\end{equation}
Eqn. ~(\ref{four}) suggests that {\it determining whether
stable equilibrium state forms following the merger of
two $1.4 M_{\odot}$ stars requires that the simulations be accurate
to better than 0.5\%.}  Considerably less 
accuracy is required as one moves away from
the maximum mass; it is only 5\% for the $0.8 M_{\odot}$ stars. Given the 
potential multiple sources of numerical error, from the initial data 
to the numerical integrations, it is by no means evident that existing 
$3+1$ numerical relativity codes can set up and track such a strong-field 
collision to 0.5\% accuracy~\cite{bride}.

The conjecture put forth in Ref~\cite{SLS}  
that when two neutron stars collide head-on from rest
at infinity, sufficient thermal pressure may be generated to support
the hot remnant in quasi-static equilibrium against collapse may be generally
correct, despite the report in Ref~\cite{MST} of a numerical
example to the contrary. The simulation cited involves
the collision of high-mass stars near the maximum mass~\cite{rad2}, 
which requires
very high numerical accuracy to simulate reliably~\cite{time}.
In the unreported case involving
lower-mass stars whose combined mass still exceeds the maximum mass, the
conjecture appears to be corroborated. The
accuracy requirements for the lower-mass case are far less stringent, and 
presumably more easily attainable. Evaluating the conjecture at high mass 
may have to await more sophisticated code development and/or larger
machines~\cite{axi}. Even with such advances, evaluating the conjecture in this
domain may still require a careful limiting proceedure based on a sequence of 
runs with increasing initial separation between the
initial stars to be confident that the initial data correctly 
corresponds to configurations which infall from rest at infinity. 

The $0.8 M_{\odot}$ simulation involves stars that have masses far below
the maximum mass. As a result, this simulation is likely to be more reliable. 
If the delayed collapse outcome found for this collision
is indeed correct, then 
we can already speculate that for neutron stars governed 
by realistic equations of state ~\cite{APR}, which currently give
maximum masses in the  range 
$1.8 M_{\odot} - 2.3 M_{\odot}$, the 
collision of two realistic (low-mass) $1.4 M_{\odot}$ stars from rest at infinity 
likely will lead to delayed collapse in accord with the conjecture.

\bigskip

\acknowledgements 

We thank T. Baumgarte, F. Lamb, M. Shibata and W.M. Suen for useful discussions.
This work has been supported in part
by  NSF Grants AST 96-18524 and PHY 99-02833 and NASA Grants NAG5-7152 and NAG5-8418
to the University of Illinois at Urbana-Champaign.


\begin{references}
\bibitem{freefall} Simulations with $10\%$ higher initial speeds were also
performed, with little change in the reported final outcome.
\bibitem{mass} For polytropes the assignment of a physical mass to
a configuration is arbitrary and has no significance, because of 
the scale freedom to choose
the gas constant $K = P/{\rho_0^{\Gamma}}$ in the equation of state
arbitrarily. Physically significant features of a polytrope
include its specific binding energy, which is scale invariant, and its 
proximity to the the maximum mass configuration, as measured by  
the the ratio of its mass to the mass of the maximum mass
configuration or by the ratio of their binding energies; 
see Ref ~\cite{SLS} and text below. Had the authors of Ref~\cite{MST}
scaled $K$ to give a more realistic maximum mass like $2.2 M_{\odot}$
instead of $1.46 M_{\odot}$,
the simulation they reported would have been for stars of mass
$2.1 M_{\odot}$, not $1.4 M_{\odot}$.
\bibitem{tol} The argument for spurious mass increase, or for spurious 
changes in the the rest mass, is similar. 

\newpage
\bibitem{rad2} A rough estimate of the gravitational radiation mass-energy
loss is $f = \Delta E/M \sim 10^{-3}$ for the $1.4M_{\odot}$ 
collision; see Ref~\cite{SLS}.
This value of $f$ is not much smaller than the maximum tolerable value 
for which an equilibrium solution and delayed collapse is 
possible ; cf. Eqn. ~(\ref{four})
\bibitem{time} It is not clear how useful it is to
attempt to distinguish, as 
in Ref~\cite{MST}, the various timescales
associated with the infall velocity, sound velocity, shock velocity,
and thermalization speed in attempting to account for the observed 
simulation outcome. To the accuracy that these timescales can be defined,
they must all be comparable on dimensional grounds, given a 
head-on collision of equilibrium stars from rest at infinity:
$t \sim R/(M/R)^{1/2}$.
\bibitem{axi} The use of adaptive gridding and/or a high-level axisymmetric
code that treats the collision in $2+1$ dimensions rather than
$3+1$ dimensions, 
like the one under development by the authors of Ref~\cite{MST},
may prove necessary.
\end{references}
\end{document}